# Intertwined chiral charge orders and topological stabilization of the light-induced state of a prototypical transition metal dichalcogenide

Yaroslav A. Gerasimenko,[1,2*] Peter Karpov,[3,4] Igor Vaskivskyi,[1,2] Serguei Brazovskii,[4,5] Dragan Mihailovic[1,2]

[1] CENN Nanocenter, Jamova 39, SI-1000, Ljubljana, Slovenia

[2] Department of Complex Matter, Jozef Stefan Institute, Jamova 39, SI-1000, Ljubljana, Slovenia

[3] Max Planck Institute for the Physics of Complex Systems, Nothnitzer Strasse 38, Dresden 01187, Germany

[4] National University of Science and Technology "MISiS", Moscow, Russia

[5] CNRS UMR 8626 LPTMS, University of Paris-Sud, University of Paris-Saclay, Orsay, France

[*] yaroslav.gerasimenko@ijs.si

**Abstract**

The fundamental idea that the constituents of interacting many body systems in complex quantum materials may self-organise into long range order under highly non-equilibrium conditions leads to the notion that entirely new and unexpected functionalities might be artificially created. However, demonstrating new emergent order in highly non-equilibrium transitions has proven surprisingly difficult. In spite of huge recent advances in experimental ultrafast time-resolved techniques, methods that average over successive transition outcomes have so far proved incapable of elucidating the emerging spatial structure. Here, using scanning tunneling microscopy, we report for the first time the charge order emerging after a *single transition outcome* in a prototypical two-dimensional dichalcogenide 1T-$TaS_2$ initiated by a *single* optical pulse. By mapping the vector field of charge displacements of the emergent state, we find surprisingly intricate, long-range, topologically non-trivial charge order in which chiral domain tiling is intertwined with unique unpaired dislocations which play a crucial role in enhancing the emergent states' remarkable stability. The discovery of the principles that lead to metastability in charge-ordered systems open the way to designing novel emergent functionalities, particularly ultrafast all-electronic non-volatile cryo-memories.

## Introduction

The search for new emergent metastable states under strongly non-equilibrium conditions with new and unexpected functionalities is currently a very popular topic in correlated complex materials [1-10]. Such states may form if the ordering timescale is shorter than the system thermalization time. Even though it is of fundamental importance to prove the principle of the existence of unique long-range order (LRO) created strongly out of equilibrium, no direct or detailed experimental evidence for this has so far been presented for any light-induced state. The problem is that the lifetimes of the transient states, usually ranging from picoseconds to microseconds [2-5, 7], limit the methods that are available [11, 12], making non-periodic mesoscopic orders and microscopic features experimentally inaccessible by even the most advanced techniques [10, 13-17].

Uniquely, the transition metal dichalcogenide 1T-$TaS_2$ (TDS) has a metastable light-induced state with a temperature-tunable lifetime, which is usefully long at low temperatures [6, 18]. This opens the possibility to investigate its emergent ordering and origins of metastability on multiple length scales and with great detail with the aid of scanning tunneling microscopy (STM).

In equilibrium, TDS hosts a variety of charge-density wave orders (Fig. 1a): a low-temperature commensurately ordered (C) ground state, or a superconducting (S) ground state, a nearly commensurate (NC) domain-textured state, a stripe state and an incommensurate (I) state. The various states can be accessed by varying the temperature [19], pressure [20, 21] or doping [22]. The NC-I non-equilibrium transition was investigated recently by electron diffraction [13, 17] and X-ray diffraction methods [14, 15] revealing relaxation dynamics on a 100 ps timescale that was interpreted to arise from domain formation and growth. However, domains were so far not directly observed, and their structure or hypothetical ordering was not reported. In all these cases so far, the final state is the same as the starting one, so strictly speaking there is no *new* emergent order.

The appearance of new hidden (H) order in TDS was revealed by the application of an ultrafast optical pulse within a narrow range of photoexcitation fluences and pulse lengths [6, 18]. Possible indication of LRO in this light-induced H state – that was claimed to be distinct from all other states of the system - came so far only very indirectly, from the linewidth and bimodal switching of the collective amplitude mode frequency observed by coherent phonon spectroscopy [6, 23]. Apart from that so far there are no data on the charge ordering structure of the state, even its ordering wavevector. The principle of non-equilibrium emergence thus remains unproven, and the underlying microscopic mechanism that leads to the remarkable metastability of the H state is poorly understood, with multiple proposed explanations [6, 24].

Recently, insulator-to-metal switching caused by direct charge injection through electrodes had been demonstrated in TDS [25-30]. Switching by charge injection through an STM tip revealed that the uniform C state can be locally broken up into non-periodic domains, which might be similar to the supercooled nearly-commensurate state [30] (Fig. 1a), raising the question if the light-induced state is the same; and if it is not, how it is different. These STM studies [29, 30] did *not* reveal any LRO of domains, so the fundamental question whether emergent metastable LRO can really form under non-equilibrium conditions remains unanswered.

A hint of new ordered states in TDS under non-equilibrium conditions was found at different photoexcitation densities in ultrafast electron diffraction experiments [31], apparently confirming that indeed new emergent states may be present on short timescales. However, the *k*-space resolution was insufficient to resolve the ordering vector or the detailed structure.

Here, using in-situ ultrafast optical switching in combination with a low-temperature high resolution STM (Fig. 1b), we show the emergence of novel intertwined domain orders in the H state that are clearly distinct from any of the other states in this material and for first time conclusively confirm emergent LRO in a non-equilibrium transition. We present a topological analysis of the nanoscale domain wall intersections, revealing new types of chiral charge ordered structures hitherto unobserved in this or any other quantum material. The experiments allow us to identify unpaired CDW dislocations and suggest a universal mechanism for achieving metastability that may be applicable to very diverse materials.

## Results

**Scanning tunneling microscopy of equilibrium and photo-induced states.** In the ground state, TDS has a uniform hexagonally ordered CDW (Fig. 2a), which is commensurate (C) with the atomic lattice. By adjusting the bias, we can resolve the atomic structure of the single CDW supercell in the C state (Fig. 2b). The characteristic truncated triangular pattern seen by STM in the sulfur layer corresponds to the $\sqrt{13}\times\sqrt{13}$ star-of-David distortion [19] in the tantalum layer (Figs. 1c, 2b), as supported by simulations (see the inset in Fig. 2b). At elevated temperatures, $T > T_{C-NC} \sim 220$ K, the uniform C order changes to a modulated nearly commensurate (NC) CDW with a highly regular array of domains, separated by smooth walls [19] (Fig. 2c).

We photoexcite fresh in-situ cleaved single crystals of TDS in the C state at 4.2 K within a UHV STM chamber with a focused 50 fs single pulse at 400 nm (see schematic in Fig. 1b). The pulse fluence was adjusted to just above the switching threshold [6], taking care to avoid heating the lattice to the NC state (see Fig. S1 for temperature estimates). The obvious effect of the optical pulse is to create domains of diverse shapes and sizes separated by sharp domain walls (DW) (Fig. 2d), to which we refer as the optically excited hidden (opt-H) state. The tiling is different every time the experiment is performed, indicating that the domain structure is determined by fluctuations rather than predetermined by sample defects or imperfections.

**Domain order on the nanoscale.** The characteristic types of domain walls that are observed are summarized in Fig. 3d. Inside the domains, both states are structurally indistinguishable from the commensurate CDW. Within the DW, the center of David stars shifts by 1 or more, not necessarily co-linear, atomic lattice vectors in different directions. The neighboring domains thus look like two C lattices shifted with respect to each other. To describe these structures, it is convenient to introduce a continuous charge displacement vector field $\mathcal{D}$ defined as the shift of the local CCDW origin, so that charge density is $\rho(\mathbf{r}) = \sum_i \cos((\mathbf{r} - \mathcal{D})Q_C^{(i)})$, $Q_C^{(i)}$ being three CDW wave vectors (Fig. 3a,b). The DW is then characterized by a direction and displacement $\delta\mathcal{D}$ between domains it separates. Owing to the David star structure (Fig. 3c), the lowest-energy displacements can occur to six nearest neighbour (NN) or six next-nearest neighbour (NNN) atomic positions within the David star, described by $\delta\mathcal{D}_{\mathrm{NN}} = \mathbf{a}_i$ or $\delta\mathcal{D}_{\mathrm{NNN}} = \mathbf{b}_i \equiv \mathbf{a}_i - \mathbf{a}_{i+1}$ respectively. They give rise to twelve possible straight domain walls, as already noted previously by Ma et al. [30].

The most common domain wall type in the opt-H state has the $\delta\mathcal{D}_{\mathrm{NNN}}$ displacement (Fig. 3d). This displacement was also found in the electrically switched state [28-30]. On the other hand, displacements $\delta\mathcal{D}_{\mathrm{NN}}$ are found to be characteristic for the NC equilibrium state, and occur only occasionally in the H state (Fig. 3d). However, the unique signature of the opt-H state is the appearance of a new type – zigzag domain walls (Fig. 3d). Unlike the straight domain walls, which are parallel to the CDW unit vectors, zigzag ones are rotated by 30° with respect to them, being parallel to $Q_C^{(i)}$ i.e. to $\mathbf{A}_C^i - \mathbf{A}_C^{i+1}$ (Fig. 3a).

On a larger scale, we can see that the opt-H state is composed of unique hexagonal structures, built with zigzag or straight NNN domain walls (Fig. 3e). The equilibrium NC state, on the other hand, has only NN-type domain structures (Fig. 3f) which are preserved also in the supercooled NC state (see Fig. S2 for analysis of the corresponding STM image from Ma et al. [30]). This highlights the difference of domain structures between the NC, electrically switched and opt-H states and dismisses the notion that the opt-H state is a supercooled NC state.

**Topological vertices.** On large scales, the domain structure in the opt-H state goes beyond the simple hexagonal arrangement, giving rise to the intersections (vertices) of up to five domain walls of different types (Fig. 4). We make their topological classification in terms of a Burgers vector construction on the CDW lattice. Its components $\mu = x, y$ are given by $\mathcal{B}_\mu = \oint \nabla\mathcal{D}_\mu \cdot d\mathbf{l}$ with integration over a contour around the vertex. If we assume that all displacements between the neighbouring domains are either NN or NNN type ($\delta\mathcal{D} \geq 2a$ is possible, but we never observed it in the experimental images), then the Burgers' vector can be reduced simply to the vector sum of the displacements inside the domain walls entering the vertex, $\mathcal{B} = \sum \delta\mathcal{D}$.

Topologically non-trivial vertices (CDW dislocations) are characterized by a non-zero Burgers vector, which in the simplest case is equal to a superlattice vector: $\mathcal{B} = A_C \equiv 3\mathbf{a}_i + \mathbf{a}_{i+1}$. This implies that no combination of the three domain walls of the same type, NN or NNN can give a non-zero Burgers vector. But vertices with three mixed DW types can be non-trivial $\mathcal{B} \neq 0$ as shown in Fig. 4b. Non-zero Burgers vectors are more readily found in vertices with four domain walls (Fig. 4d). Overall, the opt-H state is found to have a large density of dislocations in comparison with the vanishing concentration (or none) in the NC state (see Fig. S3). Remarkably, most of these dislocations are unpaired and are thus costly to remove, providing route to metastability.

In addition to the local vertex structures, there exists a long-wavelength structure which is even more intriguing (Fig. 5). Plotting the magnitude $|\mathcal{D}|$ and the angle $\alpha$ of the displacement field $\mathcal{D}$ in Fig. 5b and c respectively as a Moiré interference pattern between the C lattice and the opt-H lattice (see Fig. 5a), we observe a clear hexagonal lattice of vortices with a period $L_H \simeq 40 \pm 10$ nm (Fig. 5b-d) – which is a direct manifestation of long-range order. Repeated experiments show that the winding number $w$ [32] of the large-scale vortices is either $w = +1$ or $w = -1$, with one dominant winding direction, implying that there is a macroscopically spontaneously broken mirror symmetry in the formation of the LRO $\mathcal{D}$ vector field. For comparison, the NC state reveals much smaller vortices with period ~20 nm, as shown in Fig. 5e,f. The order in the H state that emerges after optical excitation is clearly distinguishable from other STM tip-induced states [29, 30]. In particular, in the STM-switched mosaics [30], there is no evidence of the LRO observed in the opt-H state (see Fig. S5 for a detailed analysis).

Unlike in the NC state, in the opt-H state we also observe an accompanying weak, but unambiguous modulation of the domain size whose wavelength is comparable to $L_H$ (Fig. 5g). Remarkably, topologically non-trivial vertices are concentrated in the areas with larger size of domains, whereas the areas with smaller size are free of such vertices (Fig. S3). The implication is the long-range Moiré vortex structure and the topologically protected vertices responsible for the metastability are intertwined.

**Hidden state in reciprocal space.** Fourier transforms (FTs) of the STM images allows bridging the present observations to those obtained with various diffraction techniques. To understand the relative position of the opt-H state to C, we ascertain its relation to the atomic lattice. From FT of the atomic resolution STM images (Fig. 6a,d) we can determine relative rotation angle between the CDW, $Q_H^{(i)}$, and atomic lattices. For opt-H state we obtain $\phi = 13.45° \pm 0.35°$ (error from Fig. 6d), which differs by $0.45°$ to the C state [19], and by $0.65°$ from the nearest NC state [19] (see Fig. S7 for error estimates). This value of $\phi$ correlates reasonably well with the Moiré pattern period, $2\pi/(|Q_H - Q_C|) \sim L_H$. We note that it is close to the first metastable state observed under optical excitation in time-resolved ultrafast electron diffraction data [31]. There it was tentatively associated with the triclinic CDW, but the present data shows no breaking of domains' hexagonal symmetry expected in the triclinic case [19], and thus unambiguously identifies it with the light-induced rather than any of the equilibrium states.

**Long range order**. To clarify the intricate LRO of the CDW domains in the opt-H state, we compare its reciprocal unit cell to that in NC textured state (Fig. 6b,e). The brightest peaks in FTs correspond to the CDW ordering vectors, $Q_H^{(i)}$ and $Q_{NC}^{(i)}$, while less intense satellite peaks characterise the domain structure periodicity, $k_{domain}^{(i)}$. The $Q_H^{(i)}$ peaks' angular FWHM is 0.31° (cf. Fig. 6g) is comparable to that in the nearby equilibrium states. It shows that the opt-H state order is *homogeneous* over a macroscopically large area (~200x200 nm$^2$), which is a clear evidence of LRO.

In the opt-H state the single satellite characteristic of the NC state is converted into a diffuse streak spanning a range of angles $1.7° < \Theta < 9.5°$ (Fig. 6e,f) which reflects the distribution of the domains' and shapes. Remarkably, the appearance of novel domain structures, clearly seen in the real space images, has merely a quantitative effect in the reciprocal space (see SI), making it very hard to resolve in diffraction experiments.

The cross-section *AA'* along the streak (Fig. 6h) reveals that the FT intensity is not random in *k*-space, but rather certain peaks can be identified, showing that some domain sizes are more favourable than others. This can be qualitatively understood by extending the harmonic picture developed for the NC state [33], which predicts regular periodic domain structures that minimize the elastic energy associated with the commensurability with the atomic lattice. Although the real structure is neither periodic nor regular, the attempt of fitting the satellites distribution with five harmonics gives $\phi = 13.51°$ and wavevector length $Q_H/a^* = 0.278$, consistent with the above results (see Fig. S6 for fitting and real-space reconstruction). Full description of the domains' size and shape apparently requires taking Coulomb interaction into account [34].

## Discussion

**Topological protection and metastability of the H state.** The topology of the H state is best illustrated using a graph representation. Let us represent domains by nodes, the walls that separate them by edges, and the atomic lattice by a discrete triangular mesh with a map of thirteen colours, as shown in Fig. 3c. The colours are mapped to the 13 equivalent atomic sublattices that can be occupied by the David stars' centres. The nodes occupy the knots of the triangular mesh, depending on the sublattice to which the domain belongs. Edges can have two lengths ($a$ and $a\sqrt{3}$), corresponding to NN and NNN type domain walls respectively. The trivial vertices, also classified in [35], are mapped to closed contours, and dislocations to open contours (Fig. 4).

The regular domain structures, shown in Fig. 3e,f, have nodes with the connectivity of six and with all edges of the same length. Such $Z_6$ vertexes are protected by the confinement forces along the DWs: if we want to shift the central domain to another sublattice, then all six corresponding DWs have to change to higher-energy configurations. The closed contour on the graph in Fig. 3e,f will be broken (since only $\delta\mathcal{D} < 2a$ walls are possible), leading to the creation of dislocations. Healing this mismatch requires moving also the neighboring domains to another sublattice, which is costly and makes the overall vertex conversion process very unfavorable. Geometrically, the protection originates from the different tilts of NN and NNN regular structures with respect to atomic lattice vectors (compare Fig. 3e,f). The two structures thus cannot be converted into each other continuously, and topological defects – dislocations – will appear at the connecting points (see Fig. S8 for a construction).

Dislocations seemingly play an explicit role in the metastability of the opt-H state. Indeed, their spatial map (an example is shown in the Fig. S3) shows that their significant numbers are unpaired and spatially separated. Removing them would require either finding a matching pair or pushing towards the sample boundary [36] – both processes require motion over very large distances, which is associated with a huge energy cost, and hence preventing the relaxation of the excited domain structure.

The presence of the unpaired dislocations in two dimensions precludes the establishment of the "true"' LRO [36]. Indeed, in the opt-H state the regular NNN hexagonal domain structure (Fig. 3e) has the final spatial extent and ends up with dislocations and the area of large irregular domains (Fig. S3). Therefore, the order is not suppressed globally in the system, but rather the unpaired dislocations reduce it locally. Such kind of ordering in two dimensions is commonly discussed in terms of topological transformations [36].

The overall opt-H state domain structure imaged by STM can thus be understood as a snapshot on the route of conversion of the uniform high-energy NNN regular domain structure into the lower energy uniform NN structure. Such process that happens with the creation of the dislocations can be considered as unambiguous evidence for melting of the domain structure [36-38] which generally belongs to the class of Kosterlitz-Thouless-Halperin-Nelson-Young topological transitions [36, 39].

**Dynamics of the photoinduced transition.** The equilibrium phase diagram in TDS is commonly discussed in terms of the competing effects of Fermi surface nesting [40], incommensurability strain [33] and Coulomb interactions [20, 41]. To understand why the opt-H state is different from the equilibrium states, it is necessary to consider the contributions of all these interaction that are important on different timescales during the non-equilibrium ordering process. It is known that both Mott and CDW orders start to melt on a sub-50fs timescale [42], partially restoring band dispersion [43, 44] within approximately half an oscillation period of the amplitude mode ~200 fs [43]. Photoemission experiments also reveal a rapid <50 fs transient shift of the chemical potential $\mu_c$ [44, 45], which arises from the intrinsic electron-hole imbalance of the band structure of undistorted 1T-$TaS_2$, and is tantamount to transient photodoping. Photoexcited carrier density is thus the control parameter in the formation of the H state. This also means that the Fermi surface and any possible nesting wavevectors are substantially modified while the system is out of equilibrium [46]. Coherent spectroscopy measurements of the amplitude mode during the C to H transition show that the change of low-energy electronic structure occurs within ~400 fs [23], implying that new electronic order forms during this time. The new ordering vector $Q_H^*$ is determined by nesting of the nonequilibrium Fermi surface and is affected by the Coulomb interactions between the electrons [34], whose ordering hierarchy in time is defined by their respective energy scales. However, since the new $Q_H^*$ is not commensurate with the underlying lattice, on a timescale of several picoseconds, as the electrons thermalise with the lattice, the non-equilibrium homogeneous state gains energy from breaking up into domains and charge fractionalization in the domain walls [34]. Recent experiments on the IC to NC transition show that the domain structure continues to grow on a timescale up to ~100 ps [14, 15, 17, 47]. The final result is the rotation of the transient ordering vector [31] towards $Q_H$ and the mismatch between the new H order and C order is revealed in the long-range Moiré pattern. Local changes in the domains' structure or size can be seen in the opt-H state on longer timescales, but global relaxation is prevented by the topological stabilization.

The results presented here deal with the structural aspect of the charge reordering from the C to the opt-H state. It is tempting to link the associated insulator-to-metal transition (see SI for opt-H spectrum) to the Mottness collapse caused by the structural distortion in DWs [29], however recent studies show that dimerization inside DW precludes metallization [48]. Further spectroscopic studies are thus necessary to build a coherent picture of the transition to the H state from this viewpoint.

## Conclusions

The STM measurements unequivocally lead to validation of the concept of emergence of complex LRO through microscopic many-body interactions under spatially homogeneous light excitation conditions. The emergence of long-range ordered in-plane modulation with a wavelength $\lambda_m \sim 40\ nm$ in response to a single 35 fs laser pulse is perhaps surprising. From the electronic switching timescale $\tau_H < 450\ fs$ [23], it follows that the messenger that can mediate between domains on such a timescale needs to have a velocity $v_m = \lambda_m/\tau_H \simeq 90$ nm/ps, which is comparable to the Fermi velocity of 0.1 eV electrons with an effective mass $m^*/m_e \simeq 2$. This consistency seemingly confirms the notion argued above that the emergent order originates from the quantum interference of non-equilibrium Fermi electrons.

The discovery of topological transformations that create non-trivial defects under non-equilibrium conditions revealed by these experiments represents a large step towards understanding, and whence designing new emergent metastable states in charge-ordered complex materials. TDS, with its sub-picosecond metal-to-insulator transition may lead to advances in novel non-volatile all-electronic memory devices without the involvement of either ions or magnetism through controllable switching between topologically protected electronically ordered states.

## Materials and Methods

**Sample preparation.** Crystals of $1T\text{-}TaS_2$ were synthesised by the iodine vapor transport method. The details of crystals synthesis are provided in [6]. Samples were cleaved in situ at UHV conditions and slowly cooled down to 4.2 K. The surface was characterised each time before photoexcitation to confirm a defect-free initial C state.

**Switching with ultrafast optical pulse.** The results presented here were measured in the custom-built Omicron Nanoprobe 4-probe UHV LT STM with base temperature of 4.2 K and optical access for laser photoexcitation (Fig. 1).

Photoexcitation of $1T\text{-}TaS_2$ single crystals was performed at 4.2 K (C-state) with a single 50 fs optical pulse at 400 nm (second-harmonic generation from 800 nm Ti-Sapphire laser) focused onto a ~100 um diameter spot within STM UHV chamber. This ensures highly spatially homogeneous excitation on the scale of the STM scans (~200 nm). It is important that photoexcitation energy density is carefully adjusted to be slightly above the threshold value (~0.9 $mJ/cm^2$ [6]) for switching to the hidden state. Significantly higher excitation energies result in heating of the lattice above the phase transition temperature $T_{NC\text{-}C}$, which we want to avoid.

The beam was aligned at low power to hit the apex of the STM tip. Then the tip was retracted and a high-power single pulse was applied (Fig. 1b). After approaching the tip, an area of the order of 500x500 $nm^2$ was checked for homogeneity.

**STM image correction.** To account for small errors in scanner calibrations, drift and X-Y crosstalk, raw images were corrected with a standard affine transform method that matched the observed hexagonal lattice to the ideal hexagonal lattice. In all the cases 1.174 nm was taken for CDW period, but this choice affects neither ratio between CDW and atomic periods nor the angles. The ratios were checked for consistency with positions of atomic peaks in FT whenever possible. The ratios were checked for all – H, C, NC – states. Contrast adjustment in Fourier transforms is routinely used for analysis of STM data in $1T\text{-}TaS_2$ and is described thoroughly in literature [19]. Here all FT images has max/min ratio of ~2.5, approximately an order of magnitude of absolute FT amplitude. Only parts of the images used for FT are shown in Fig. 2. Full-scale versions of those for the hidden state are shown in Fig. S9.

**Acknowledgments**

We thank Petra Sutar for providing the 1T-$TaS_2$ crystals used in the present studies. We are grateful to A. Kranjec for his help in domain recognition. This work was supported by European Research Council Advanced Grant ERC-2012-ADG-20120216 "Trajectory". S.B. and P.K. acknowledge the financial support from the Ministry of Science and Higher Education of the Russian Federation in the framework of Increase Competitiveness Program of NUST "MISiS" No. K2-2017-085. P.K. acknowledges the support of Alexander von Humboldt Foundation.

**Author contributions:** Y.G. and I.V. performed the measurements; Y.G., P.K., and D.M. carried out the analysis. Y.G. and D.M. wrote the paper. S.B. and P.K. contributed to the theoretical interpretation. All authors discussed the results and contributed to the manuscript preparation.

## Figures

**Fig. 1. Charge ordering of equilibrium and non-equilibrium states of 1T-$TaS_2$ and a schematic diagram of the experiment.** (**a**) Schematic charge ordering in the equilibrium states: the incommensurate (IC), nearly commensurate (NC), triclinic (T) and commensurate (C) states observed upon heating [19]. Electron-rich (poor) areas are shown with red (blue) color. (**b**) Schematic view of the combined STM and SEM system with optical access for ultrafast laser pulse excitation. The SEM image shows a cleaved ***a-b*** surface of TDS single crystal in contact with the sharp tungsten STM tip.

After beam alignment the tip is retracted and the area underneath is illuminated with a focused single ultrafast laser pulse. (**c**) The David star polaron structure in the C state on the 1T-$TaS_2$ atomic lattice, where twelve atoms are displaced towards the central Ta with the one extra electron, as indicated by the arrows.

**Fig. 2. Raw STM images of the hidden and equilibrium CDW states in 1T-$TaS_2$.** (**a**) The large-area STM image ($V_t$ = –800 mV, I = 1.5 nA) of the commensurate CDW at 4.2 K. Bright dots correspond to the David stars. (**b**) The atomic resolution STM image ($V_t$ = 100 mV, I = 200 pA) of the CCDW, shows the reconstruction in the sulfur layer caused by the star of David (red) distortion in the tantalum layer. Inset shows the simulated pattern. (**c**) The large-area STM image of the nearly commensurate CDW state at 300 K ($V_t$ = –250 mV, I = 150 pA). (**d**) The extensive-area STM image ($V_t$ = –200 mV, I = 100 pA) of the opt-H CDW state at 4.2 K after switching with a single ultrafast optical pulse (see Fig. S9 for full image).

**Fig. 3. New domain walls and domain structures in the opt-H state.** (**a**) Schematic structure and vector definitions in the CCDW: grey – atomic lattice, black – David stars. (**b**) Definition of the displacement vector $\mathcal{D}$ – the change in local origin of CCDW. To avoid unphysical jumps, we identify $\mathcal{D} \equiv \mathcal{D} + \mathbf{A}_C$. Atomic lattice color map shows with different colors thirteen equivalent sublattices that can be occupied by the David star center. We set $\mathcal{D} \equiv 0$ for the red sublattice. (**c**) A graph representation of a NNN domain wall (panel d) on the atomic lattice. (**d**) Three types of domain walls in the opt-H state classified by the displacement vector. NNN and zig-zag walls are unique to the H state. (**e**) Regular domain structure exclusive to the opt-H state and its equivalent graph representation. All walls (graph edges) are of the same NNN type. (**f**) Regular domain structure of the equilibrium NC state. All walls (graph edges) are of NN type.

**Fig. 4. Topological vertices in the H state.** Topologically trivial (**a**, **c**, **e**) and non-trivial (**b**, **d**) vertices observed in the opt-H state, and their respective graph representations. The closed contours on the graphs (a, c, e) have the Burgers vector $\mathcal{B} = 0$. Open contours which start and end at the different nodes of the same color have $\mathcal{B} = \mathbf{A}_C$ and thus contain a dislocation.

**Fig. 5. Long-range vortex order in the H and NC states.** (**a**) The definition of the displacement field magnitude $|\mathcal{D}|$ and its direction $\alpha$ w.r.t. the C lattice. We calculate the above with respect to the Wigner-Seitz cell of the superlattice and origin in some arbitrary domain. (**b, c**) Large scale real space images of the opt-H state showing $|\mathcal{D}|$ and $\alpha$ for each David star with respect to its position in the initial C lattice. The image exhibits long-range hexagonal vortex lattice on a scale of ~40 $\pm$ 10 nm. The vortices (winding number –1) are indicated by white circular arrows. (**d**) The Moiré pattern between $Q_H$ and $Q_C$ lattices calculated on the same grid. The vortices of the misfit vector field are shown by white circular arrows. A cartoon in the bottom right corner schematically shows the actually observed hexagonal tessellation of domains: instead of a continuous change, $\mathcal{D}$, (shown by the black arrows) is fixed inside a domain and changes on crossing a domain wall. (**e, f**) Real space images of the NC state showing the misfit amplitude (**e**) and the direction (**f**) on the same scale as (b) and (c). Both tiling and period are different than in the opt-H state. (**g**) The modulation of the domain size in the opt-H state on a scale of ~50 nm, emphasized by shading (the cut position is shown in Fig. S9).

**Fig. 6. Reciprocal space analysis of the opt-H state in relation to equilibrium states.** (**a**) A Fourier transform (FT) of the atomically resolved STM image of C state (Fig. 2b) shows atomic (black) and CDW (red) reciprocal cells and angle $\phi$ between them. $a^*$ and $b^*$ are reciprocal atomic vectors and $Q_C^{(i)}, (i = 1, 2, 3)$ – the CDW ones. (**b**) FT of the STM image of NC state (Fig. 2c). (**c**) Zoom in panel (b) shows the highest-intensity fundamental peak (red), surrounded by 3 satellites (blue). The domain structure vector is $k_{domain}^{(2)} = Q_{NC}^{(2)} - Q_{sat}^{(2)}$ (**d**) FT of the atomic resolution STM image of the opt-H state

(see Fig. S9). (**e**) FT of the large-scale STM image of the opt-H state (Fig. 2d, FT was calculated for the full image in Fig. S9) shows a CDW cell (red). The satellite streak spread is $\Theta \approx 9.5°$. (**f**) Zoom in the black box in panel (e). (**g**) Azimuthal cross-section data of the fundamental peaks in opt-H ($Q_H^{(1)}$), C ($Q_C$, T = 4.2 K) and NC ($Q_{NC}$, T = 205 K, third-order peak) states. The lines are gaussian fits. FWHM = 0.31° for the opt-H state. (**h**) *AA'* cross-section of the streak in panel (f) with a 5-pixel lateral averaging. The arrow shows the n = 1 CDW harmonic.

**Figure 1**

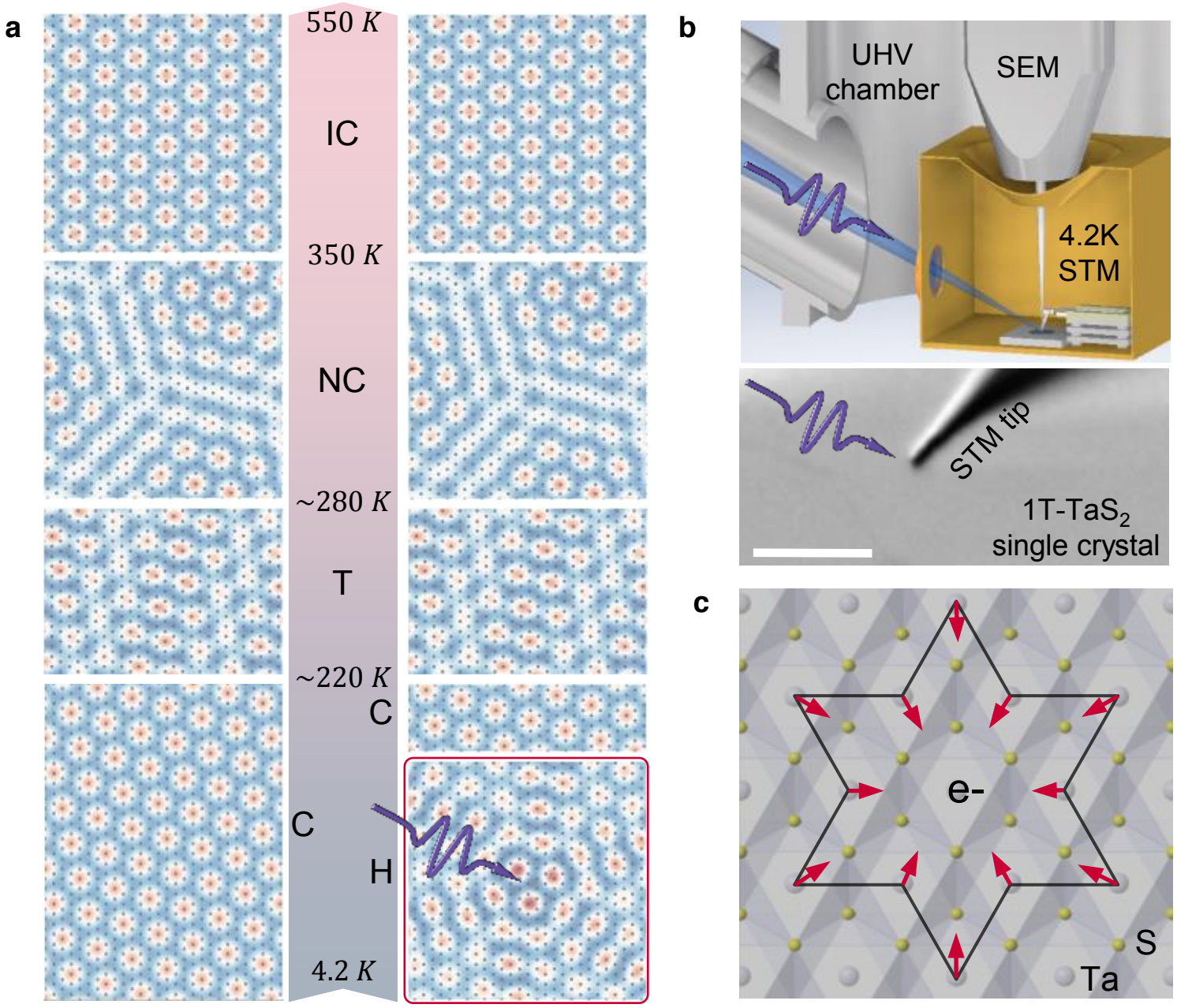
a
550 K
IC
350 K
NC
~280 K
T
~220 K
C
C
H
4.2 K
b
UHV
chamber
SEM
4.2K
STM
STM tip
1T-$TaS_2$
single crystal
c
e-
S
Ta

Figure 2

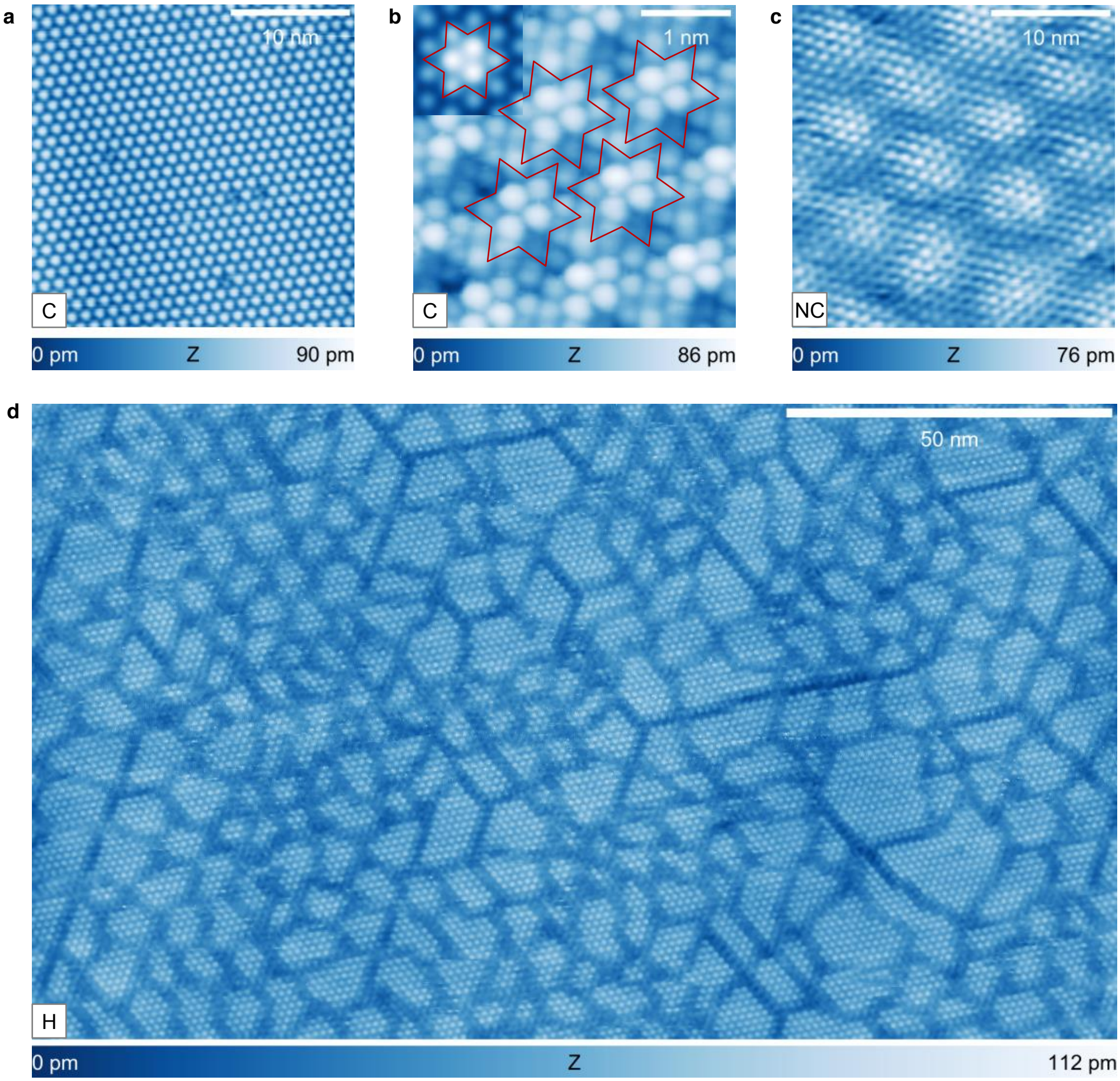

a
10 nm
C
0 pm
Z
90 pm
b
1 nm
C
0 pm
Z
86 pm
c
10 nm
NC
0 pm
Z
76 pm
d
50 nm
H
0 pm
Z
112 pm

**Figure 3**

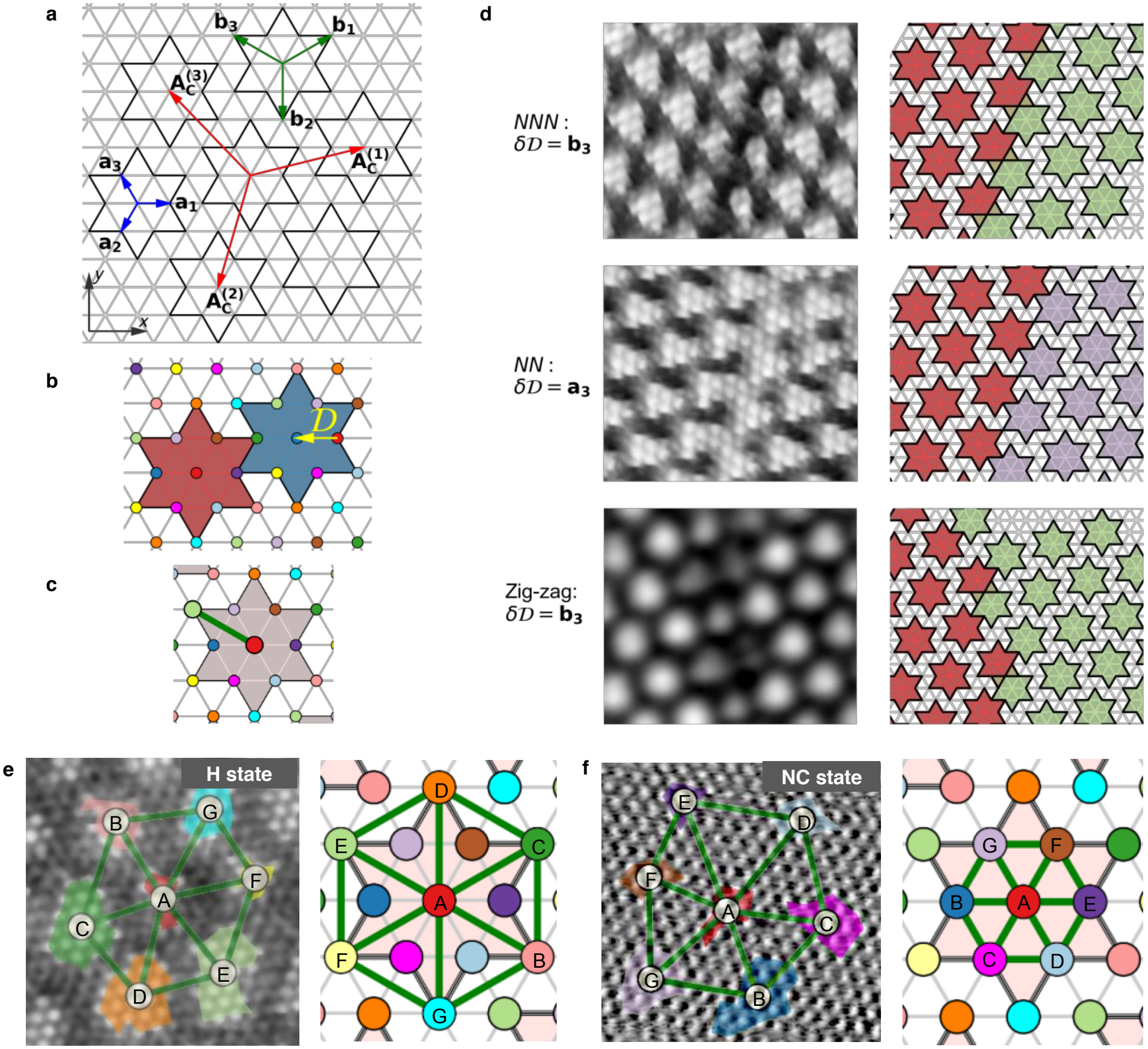

a
b
c
d
e
f
$\mathbf{b}_3$
$\mathbf{b}_1$
$\mathbf{b}_2$
$\mathbf{A}_C^{(3)}$
$\mathbf{A}_C^{(1)}$
$\mathbf{A}_C^{(2)}$
$\mathbf{a}_3$
$\mathbf{a}_1$
$\mathbf{a}_2$
y
x
$D$
$NNN$ :
$\delta D = \mathbf{b}_3$
$NN$ :
$\delta D = \mathbf{a}_3$
Zig-zag:
$\delta D = \mathbf{b}_3$
H state
NC state
A
B
C
D
E
F
G

**Figure 4**

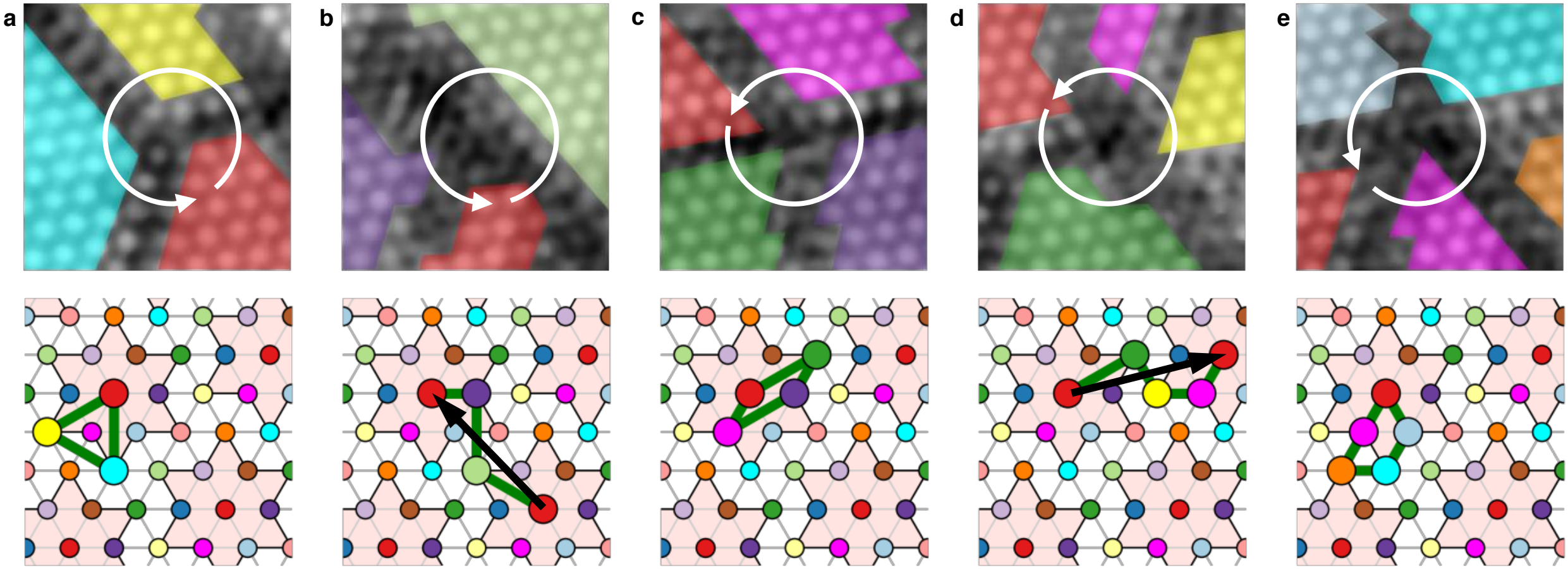

**Figure 5**

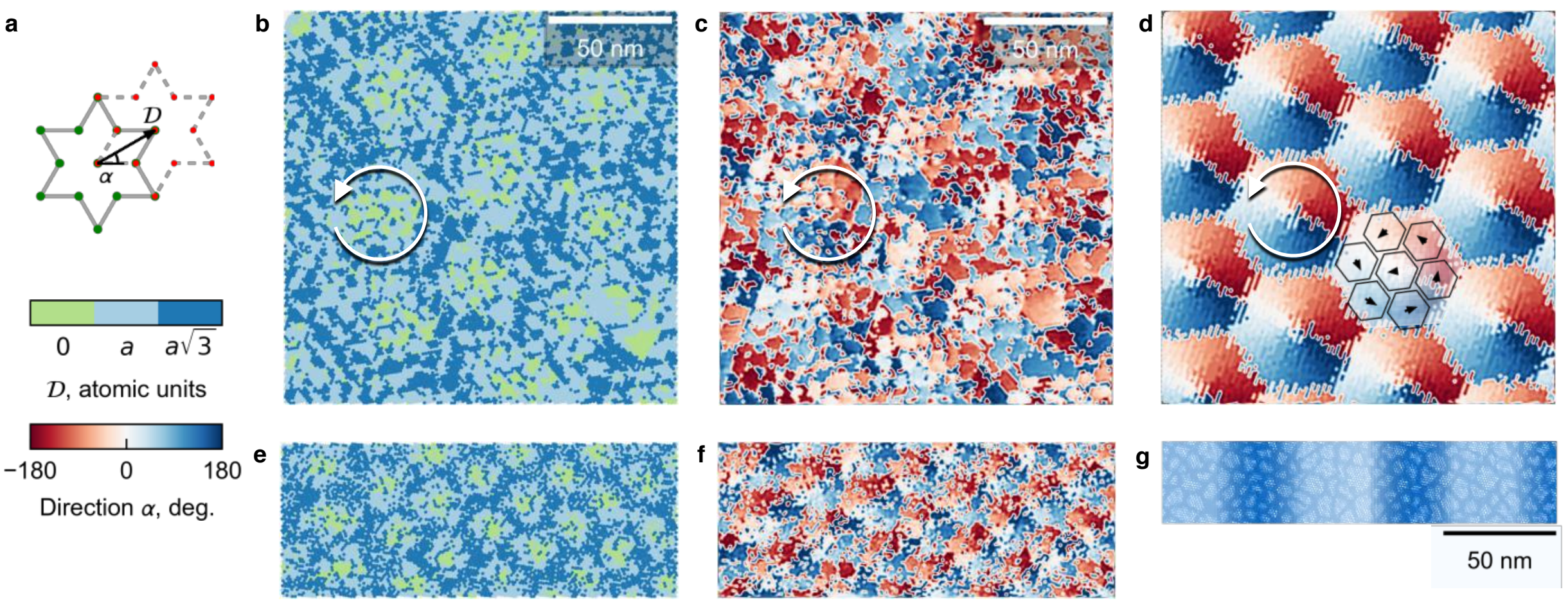

a
b
c
d
e
f
g
50 nm
50 nm
50 nm
$\mathcal{D}$
$\alpha$
0
a
$a\sqrt{3}$
$\mathcal{D}$, atomic units
−180
0
180
Direction $\alpha$, deg.

**Figure 6**

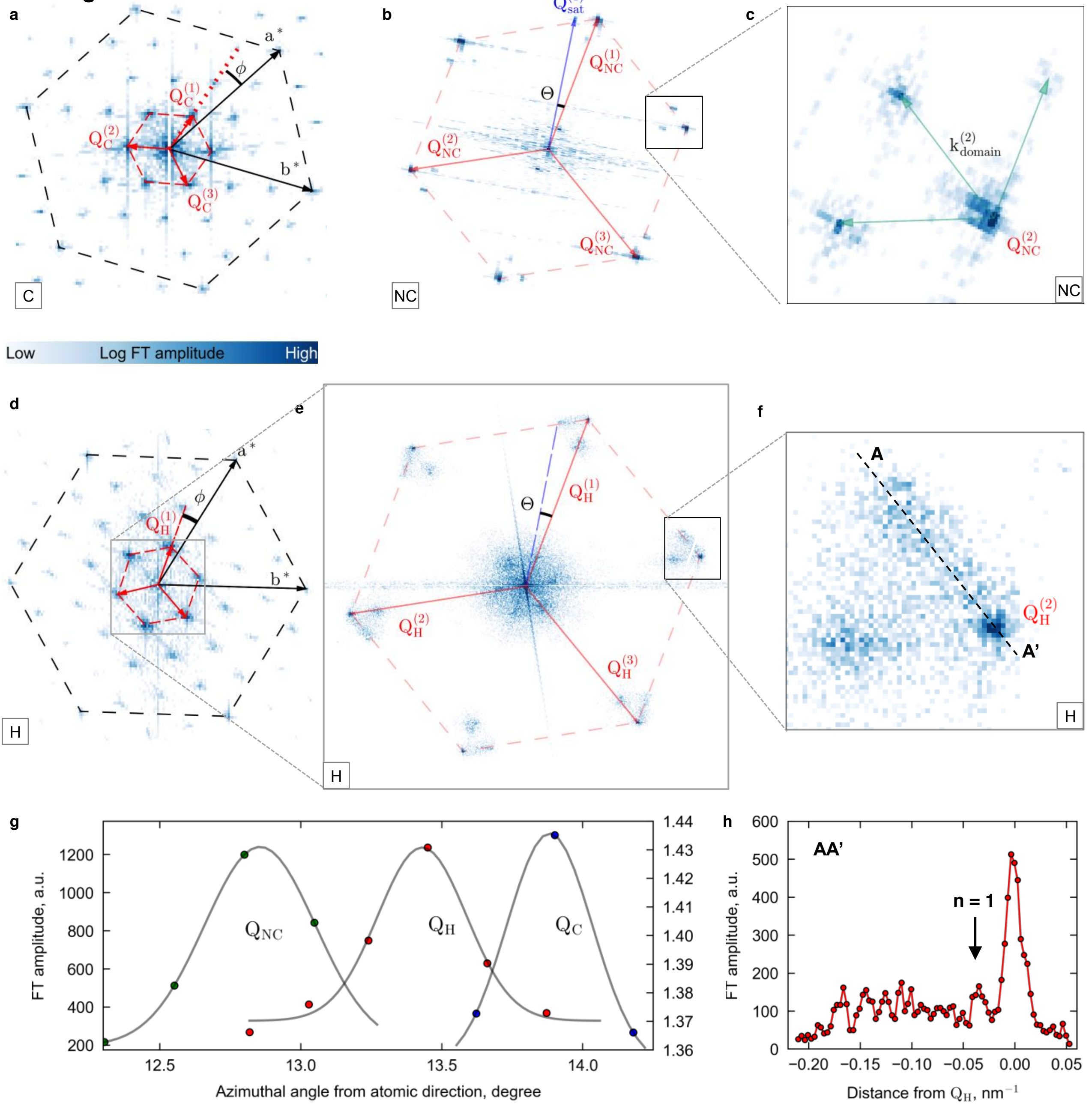

a
b
c
d
e
f
g
h
C
NC
NC
H
H
H
Low
Log FT amplitude
High
AA'
n = 1
A
A'
FT amplitude, a.u.
Azimuthal angle from atomic direction, degree
Distance from Q_H, nm^-1